\documentclass{hpc-ua}
\usepackage{tabularx}
\usepackage{graphicx}

\begin{document}

\selectlanguage{english}

\title{From Quantity to Quality: Massive Molecular Dynamics Simulation of Nanostructures under Plastic Deformation in Desktop and Service Grid Distributed Computing Infrastructure}

\author{Olexander Gatsenko\affiliationmark{1}, Lev Bekenev\affiliationmark{1}, Evgen Pavlov\affiliationmark{2}, Yuri G. Gordienko\affiliationmark{1}\affiliationmark{*}}

\affiliation{%
\affiliationmark{1}G.V.Kurdyumov Institute for Metal Physics, National Academy of Sciences of Ukraine, Kiev, Ukraine \\
\affiliationmark{2}Taras Shevchenko National University of Kiev, Kiev, Ukraine}
    \email{\affiliationmark{*}gord@imp.kiev.ua}

\maketitle

\begin{abstract}
The distributed computing infrastructure (DCI) on the basis of BOINC and EDGeS-bridge technologies for high-performance distributed computing is used for porting the sequential molecular dynamics (MD) application to its parallel version for DCI with Desktop Grids (DGs) and Service Grids (SGs).
The actual metrics of the working DG-SG DCI were measured, and the normal distribution of host performances, and signs of log-normal distributions of other characteristics (CPUs, RAM, and HDD per host) were found. The practical feasibility and high efficiency of the MD simulations on the basis of DG-SG DCI were demonstrated during the experiment with the massive MD simulations for the large quantity of aluminum nanocrystals ($\sim10^2$-$10^3$). Statistical analysis (Kolmogorov-Smirnov test, moment analysis, and bootstrapping analysis) of the defect density distribution over the ensemble of nanocrystals had shown that change of plastic deformation mode is followed by the qualitative change of defect density distribution type over ensemble of nanocrystals. Some limitations (fluctuating performance, unpredictable availability of resources, etc.) of the typical DG-SG DCI were outlined,
and some advantages (high efficiency, high speedup, and low cost) were demonstrated.
Deploying on DG DCI allows to get new scientific \emph{quality} from the simulated \emph{quantity} of numerous configurations by harnessing sufficient computational power to undertake MD simulations in a wider range of physical parameters (configurations) in a much shorter timeframe.

\end{abstract}

\section{Introduction}
Search for new nanoscale functional devices is considered as ``El Dorado'' and stimulates ``Gold Rush'' in the modern materials science. But controlled fabrication of nanoscale functional devices requires careful selection and tuning the critical parameters (elements, interaction potentials, regimes of external influence, temperature, etc) of atomic self-organization in designed patterns and structures for nanoscale functional devices. That is why molecular dynamics (MD) simulations of nanofabrication processes with physical parameter decomposition for parameter sweeping in a brute force manner are very promising. Usually MD-simulations for realistic configurations take huge resources of supercomputers with large shared memory and big number of CPUs. But the recent advances in computing hardware, algorithms, and infrastructures, especially in development of distributed computing infrastructures, allow us to elaborate the efficient methods for solving these tasks without expensive scaling-up. For example, the distributed computing model on the basis of the
BOINC \cite{Kacsuk2009},
XtremWeb-HEP \cite{cappello2005computing},
OurGrid \cite{cirne2006labs},
EDGeS \cite{urbah2009edges}
platforms for high-performance distributed computing becomes very popular due to feasibility to use donated computing resources of idle PCs and integration with global computing grid. The main aims of the work are as follows:
\begin{itemize}
  \item to analyze the actual metrics of the working DG-SG DCI for the better understanding of the nominal and available computing resources and better planning the large scale MD simulations in the variable DCI,
  \item to demonstrate the practical feasibility and high efficiency of the MD simulations on the basis of DG-SG DCI.
\end{itemize}

\section{Distributed computing infrastructure}
\begin{figure}[!ht]
        \begin{center}
            \includegraphics[width=10cm,angle=0]{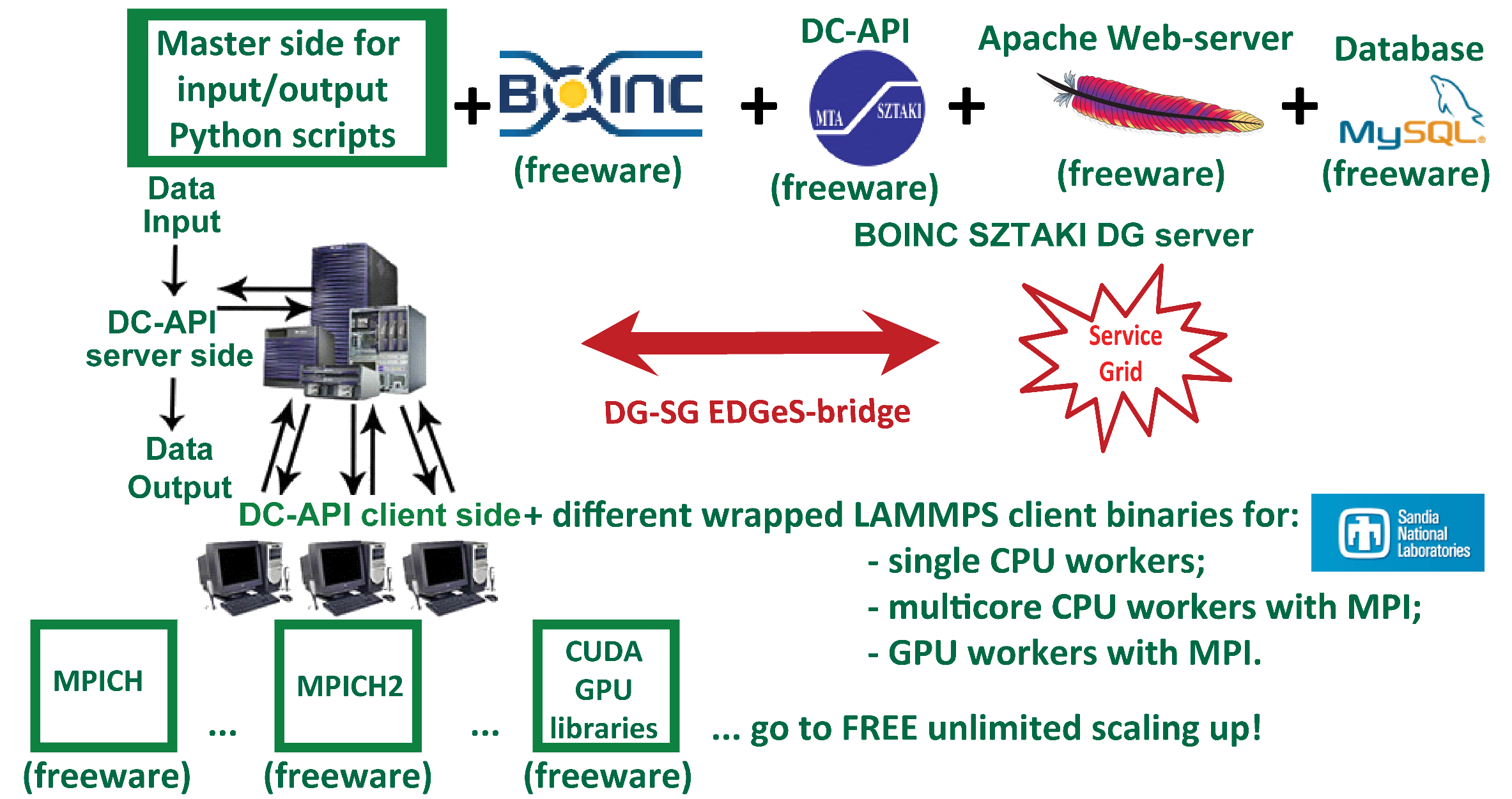}
        \end{center}
        \caption{The general scheme of porting LAMMPSoverDCI application to DG-SG DCI for massive MD simulations with job distribution by SZTAKI DG approach \cite{Kacsuk2009}.}\label{Fig01_Porting}
    \end{figure}

\begin{figure}[!ht]
        \begin{center}
            \includegraphics[width=9cm,angle=0]{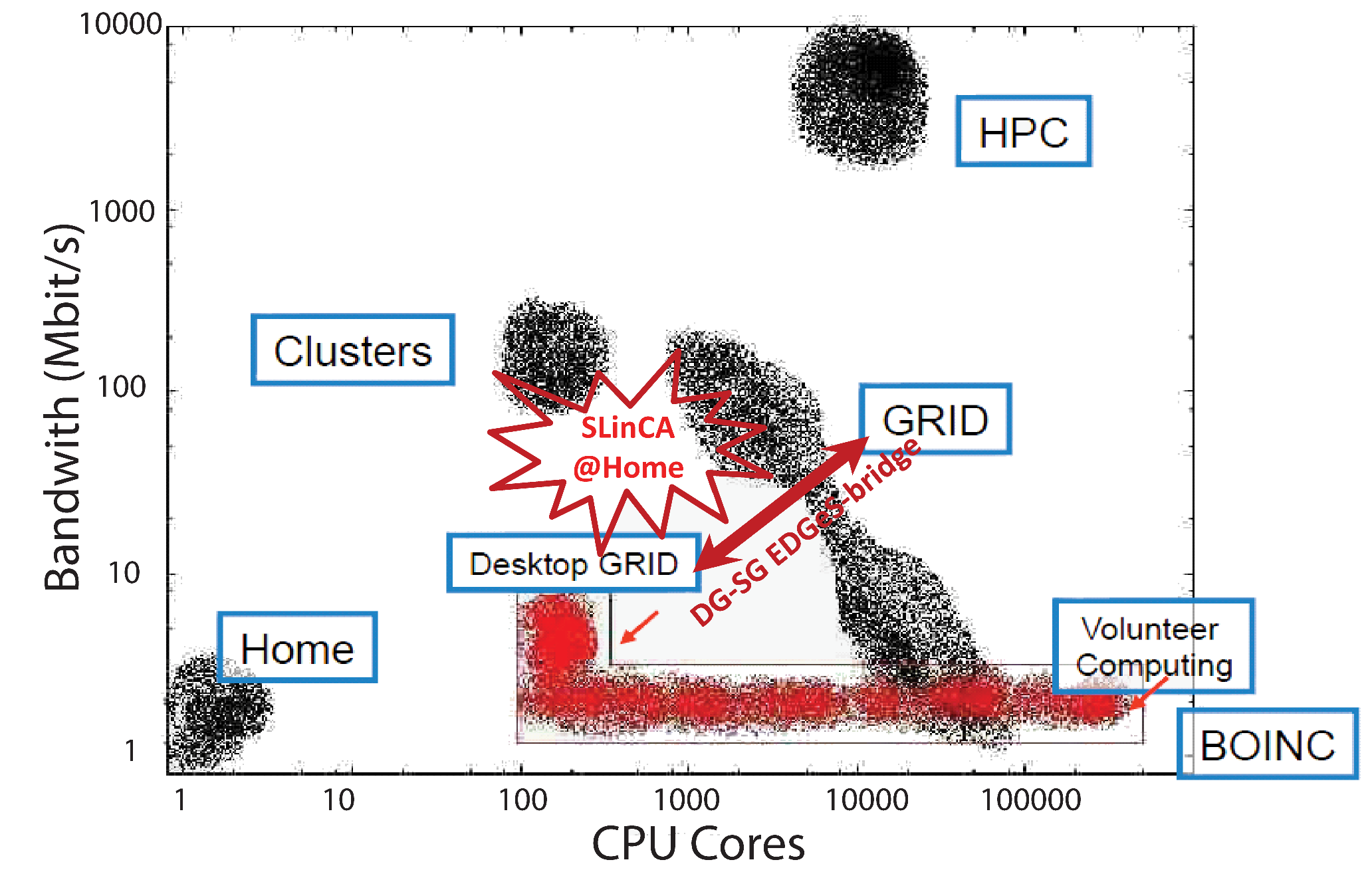}
        \end{center}
        \caption{DG-SG DCI \emph{SLinCA@Home} used for massive MD simulations in the context of the available HPC approaches (adaptation of the original figure by courtesy of Fermin Serrano).}\label{Fig02_Diagram}
    \end{figure}

Many sequential applications by slight modifications in its code could to be ported to the parallel version for worker nodes of a distributed computing environment (DCI) as Desktop Grid (DG) by means of the BOINC software platform and availability of simple and intuitive Distributed Computing Application Programming Interface (DC-API) \cite{Kacsuk2009}.
For this purpose the very popular non-commercial open-source package LAMMPS by Sandia Labs
\cite{plimpton1995fast}
(\texttt{http://lammps.sandia.gov}) was selected as a candidate for porting to DG DCI as the DG-enabled application \emph{LAMMPSoverDCI}
(Fig.~\ref{Fig01_Porting}) (the details of such porting were given recently in \cite{4GLCGW10}).
The typical simulation of the investigated nanostructure under 1 configuration of physical parameters --- for instance, metal single crystal with $10^7$ atoms with embedded atom potentials for 1-10 picoseconds of the simulated physical process --- takes approximately 1-7 days on a single modern CPU.
But by means of the new technology of high-performance computations (HPC) on the basis of DG-SG DCI the massive MD simulations of plastic deformation processes can be carried out for the large quantity of Al nanocrystals ($10^2$-$10^3$).
These MD simulations were conducted in several powerful computing environments:
in the created DG-SG DCI \emph{SLinCA@Home} (\texttt{http://dg.imp.kiev.ua}), which is the most powerful scientific DCI in Ukraine
\cite{GatsenkoCGW09},
and with involvement of computational resources of the European SG of the European Grid Initiative (EGI)
by means of EDGeS-Bridge technology
\cite{urbah2009edges}.
Deploying \emph{LAMMPSoverDCI} application to this DG-SG DCI potentially allows to utilize thousands of machines (Fig.~\ref{Fig02_Diagram})
for simulations of numerous initial configurations at the same time,
but the actual performance can be very variable due to unstable availability of distributed resources.

\section{Results of MD computer simulations in the context of computer science}

The quantitative analysis of the available and actually used resources in DG-SG DCI \emph{SLinCA@Home} was carried out to estimate feasibility of DG-SG DCI for MD simulations in materials science. Its general characteristics were calculated for the typical usage during the MD-simulations of plastic deformation of many nanocrystals.

\subsection{The nominal and actual characteristics of the DG-SG DCI}

The whole number of the registered hosts in the DG-SG DCI used for MD simulations was 4161 at the moment of writing this paper, including 649 hosts actually worked for all (more than 30 versions) applications installed in the DG-SG DCI, and 189 hosts used for the LAMMPS-application in the current example, which is described below.

It should be emphasized that the actual characteristics of DG-SG DCI are not uniform like in a cluster or Service Grid infrastructure. Like any other DCI with variable resources, the computing performance of DG-SG DCI \emph{SLinCA@Home} cannot be constant, but fluctuates with time depending on the current state of the donated resources: some hosts can be inactive or unplugged, and, in contrary, some strong hosts can be attached or started unpredictably. Some long jobs can be obtained by hosts and returned much later that they were finished actually. That is why, the high nominal characteristics of the such DCI, like many projects in volunteer computing niche, should not be overestimated and their actual averaged performance should be estimated for the better planning the massive simulations with many long jobs.

\begin{figure}[!ht]
        \begin{center}
            \includegraphics[width=6cm,angle=0]{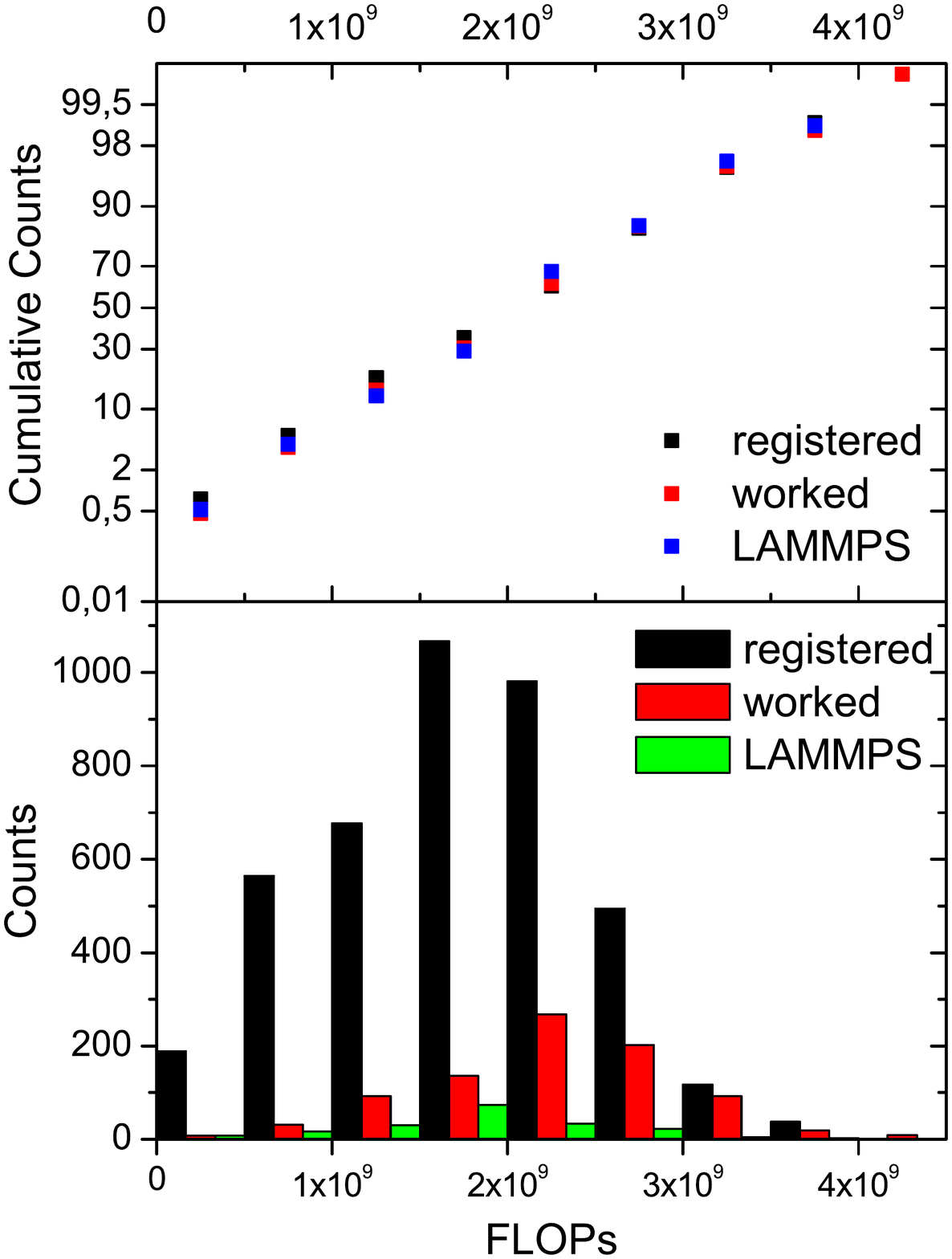}
            \includegraphics[width=6cm,angle=0]{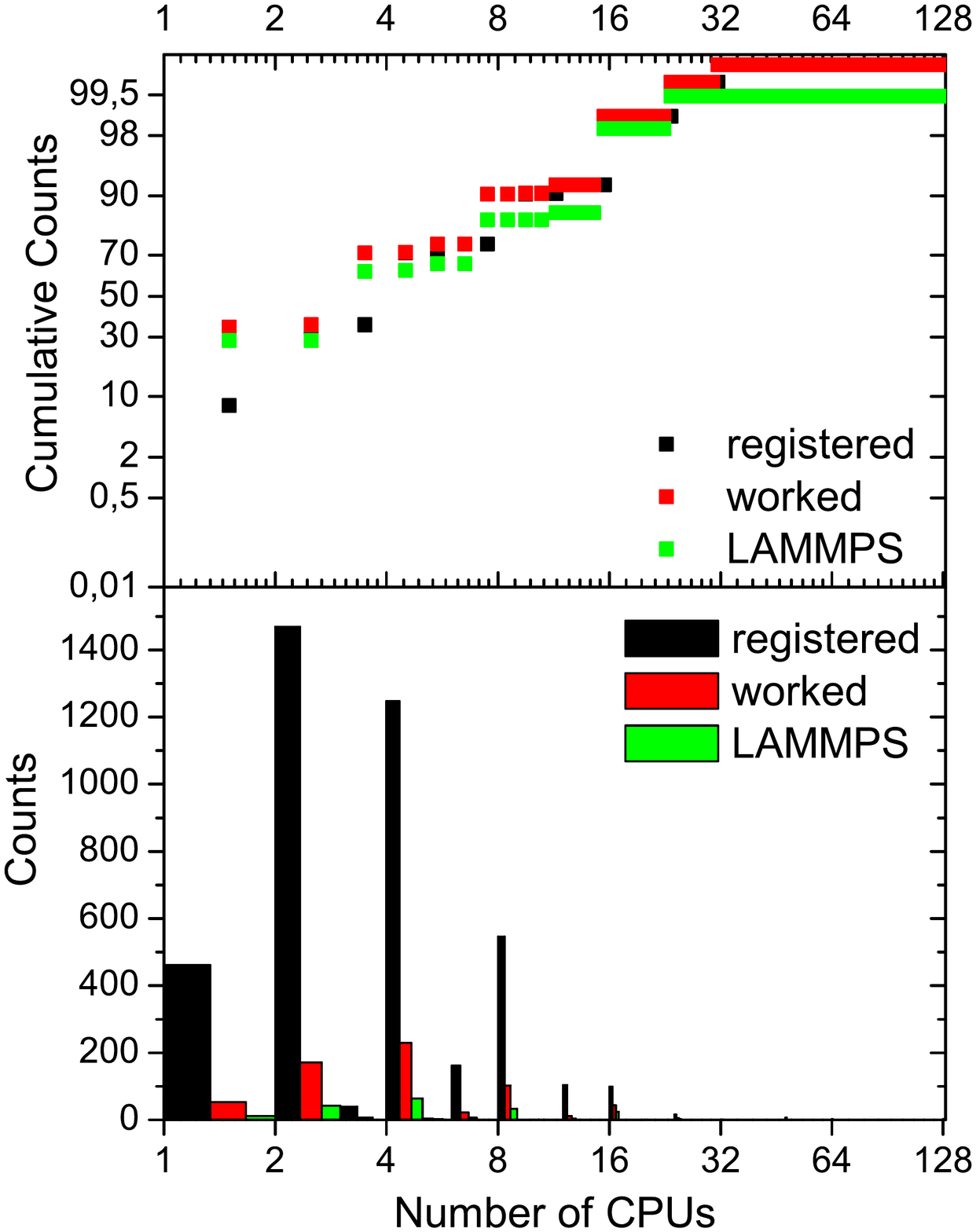}
        \end{center}
        \caption{Distribution of performances in floating point operations per second (FLOPs) in linear-probabilistic coordinates (left) and the number of CPUs per host in the logarithmic-probabilistic coordinates (right) among the registered hosts (black), the worked hosts (red), and the hosts used for LAMMPS-application (green).}\label{Fig03_CPU_FLOPs}
    \end{figure}

DG-SG DCI \emph{SLinCA@Home} demonstrates the wide variety of the numbers of CPUs, for example, from 1 CPU in the usual domestic PCs to 128 CPUs in the Service Grid connected to DG-SG DCI \emph{SLinCA@Home}. To estimate its registered and actual potential the analysis of the nominal (registered hosts) and actually used (worked hosts) computing resources was carried out on the basis of the data queries to the DCI database (mysql) with the history of calculations for the selected LAMMPS-application. The distribution of the host performances in floating point operations per second (FLOPs) is very wide also and can be approximated by the normal distribution (Fig.~\ref{Fig03_CPU_FLOPs},left). The number of CPUs per host (Fig.~\ref{Fig03_CPU_FLOPs},right) does not follow the normal distribution and can be better approximated by log-normal distribution for the low numbers of CPUs. The straight lines of the cumulative distribution function in linear-probabilistic coordinates in the top part (Fig.~\ref{Fig03_CPU_FLOPs},left) and logarithmic-probabilistic coordinates in the top part (Fig.~\ref{Fig03_CPU_FLOPs},right) give the visual evidences for these approximations.

The first assumption as to the normal distribution of the host performances (in FLOPs) can be explained by the additive growth of the CPU frequency in the modern machines. The stochastic attachment/detachment of the hosts to the DG-SG DCI can be described by evolution with diffusive law that gives rise to the normal distribution. The more detail investigation of such aggregation kinetics in the more general case was given elsewhere \cite{gordienko2011generalizedIJMPB}.
The last assumption as to the log-normal distribution of the number of CPUs per host can be explained by the multiplicative growth (with the binary multiplicator) of the number of cores in the modern machines (going mostly in the sequence 1, 2, 4, 8, ... with rare exceptions, like 6 or 12 cores). That is why, the stochastic attachment/detachment of the hosts to the DG-SG DCI can be described by evolution with Gibrat's rule of proportionate (multiplicative) growth that gives rise to the log-normal distribution \cite{gibrat1931les}. The more detail investigation of these properties of the computing resources in the typical DG-SG DCI will be given elsewhere \cite{gordienko2012multiplicative}.

\begin{figure}[!ht]
        \begin{center}
            \includegraphics[width=6cm,angle=0]{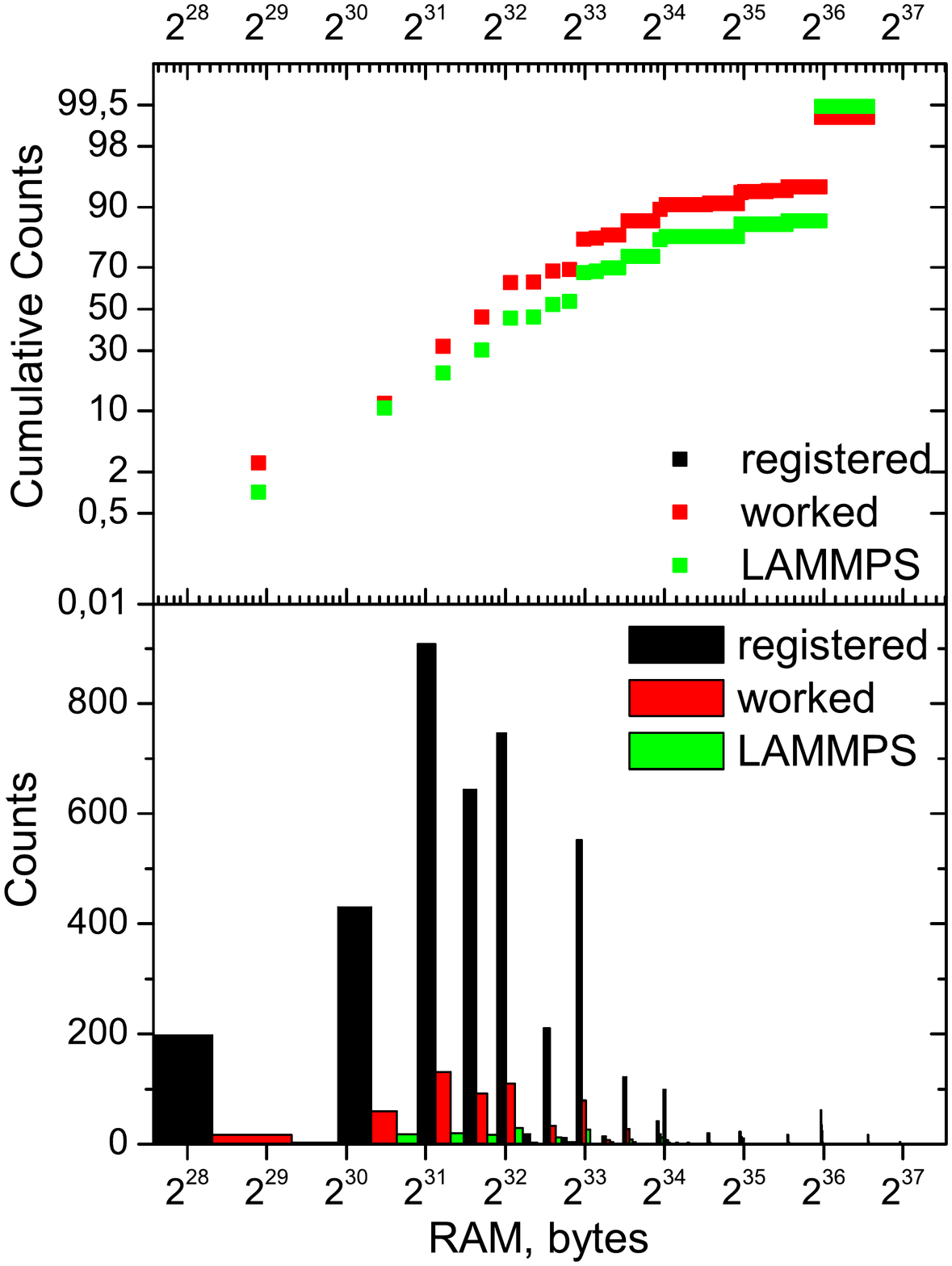}
            \includegraphics[width=6cm,angle=0]{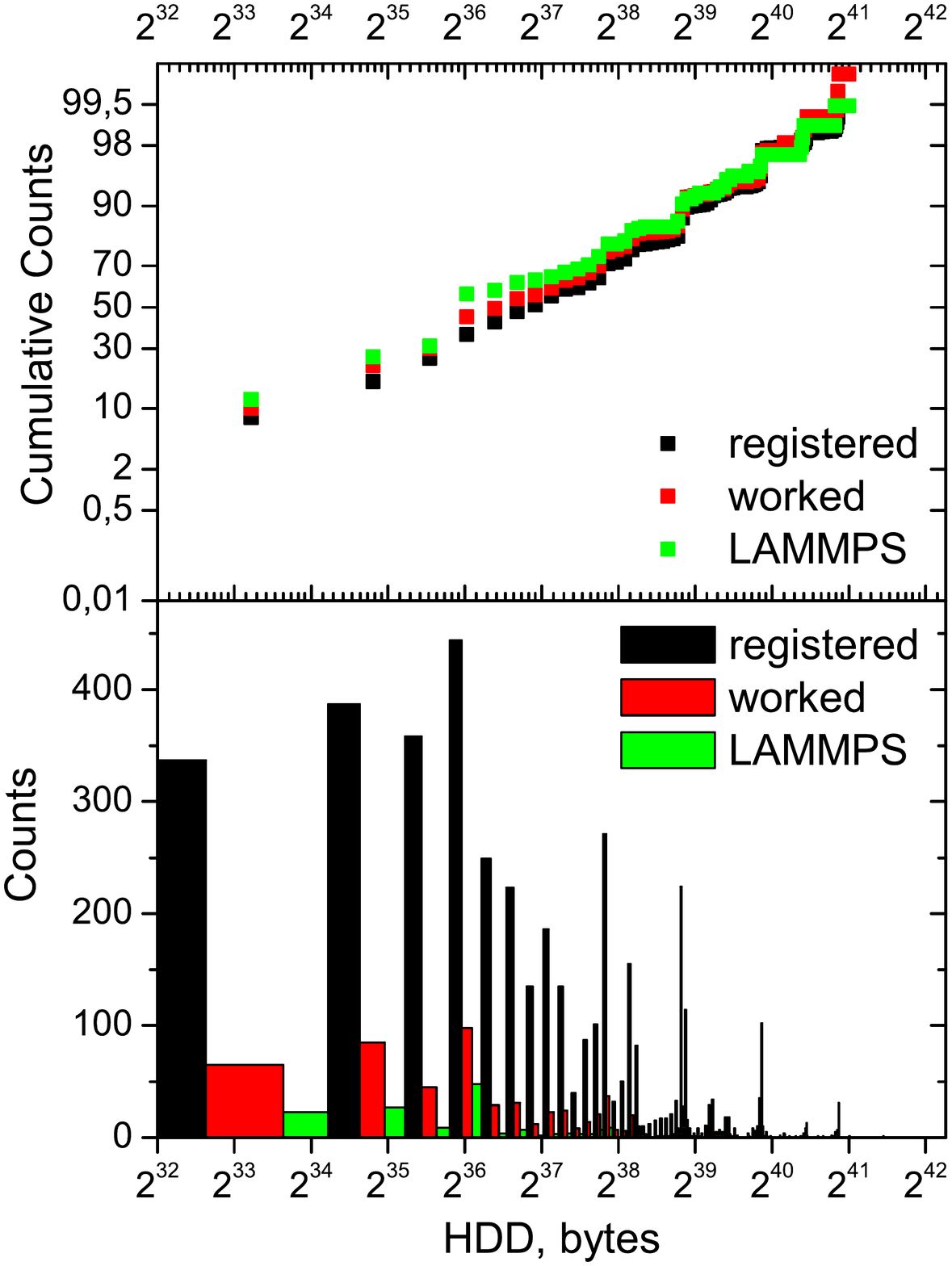}
        \end{center}
        \caption{Distribution of the RAM sizes (left) and HDD sizes (right) for the registered hosts (black), the worked hosts (red), and the hosts used for LAMMPS-application (green).}\label{Fig04_RAM_HDD}
    \end{figure}

The similar considerations can be applied for the distributions of the sizes of random-access memory (RAM) (Fig.~\ref{Fig04_RAM_HDD},left) and sizes of hard disk drives (HDD) (Fig.~\ref{Fig04_RAM_HDD},right), those do not follow the normal distribution and can be better approximated by log-normal distribution for the not very big values of RAM and HDD sizes (the left tails and central parts of these distributions). The bad approximation for the big values of RAM and HDD sizes (the right tails of these distributions) can be explained by the attachment of the big SG-hosts or clusters, which performances (FLOPs, CPUs, RAM, HDD) evolve not stochastically, but abnormally fast and deterministically (by the strategic plans of their corporative owners).

It should be noted that the more strict statistic methods for estimation of relation of these empirical distributions to the proposed distributions (normal and log-normal) will be necessary, but it is not possible at this stage, because of the discrete nature of the available values of some parameters. For example, the number of CPUs per host can have only the limited set of values (1, 2, 4, 6, 8, 16, 32, 48, 64, 128) and the number of hosts with more than 8 CPUs is very limited (that creates the long right tail of the distribution). This observation is applicable for RAM and HDD distributions also (Fig.~\ref{Fig04_RAM_HDD}). The additional aspect is that the set of hosts (the worked hosts and the hosts used for LAMMPS) is artificially limited by each type of application to have enough computing resources to run the task (the certain type of jobs), and these limitations are different for different tasks. That is why the different and varying with time subsets of all hosts were used and analyzed in the experiment.

Nevertheless, some averaged values for the nominal and actual characteristics of hosts in DG-SG DCI \emph{SLinCA@Home} (the mean values with their standard deviations) for the three sets of hosts (registered, worked, and used in the experiment with LAMMPS application) can be estimated and they are given in the Table~\ref{Table01}. The growth of the mean values for the subsets of hosts ``Worked'' and ``Used in LAMMPS'' is explained by several limitations (namely, by the minimal RAM and job runtime) imposed on the different tasks. It should be noted that the standard deviation values for all parameters (except for FLOPs) are higher than their mean values, that is typical for asymmetric long-tailed distributions like log-normal ones.

\begin{table}[!ht]
  \centering
    \caption{The averaged values (i.e. mean$\pm$standard deviation) of the nominal and actual characteristics of hosts in DG-SG DCI \emph{SLinCA@Home}.}
    \label{Table01}
    \begin{tabular}{|c|c|c|c|c|}
      \hline
                    & CPUs per host  & GFLOPs           & RAM (GB)       & HDD (GB) \\ \hline
      Registered    & $4.30\pm4.95$  & $2.25\pm0.76$ & $6.68\pm12.15$ & $257\pm371$ \\ \hline
      Worked        & $5.3\pm6.5$  & $2.3\pm0.7$ & $10\pm18$ & $220\pm310$ \\ \hline
      Used in LAMMPS & $6.7\pm10$  & $2.3\pm0.7$ & $16\pm22$ & $210\pm320$ \\
      \hline
    \end{tabular}
\end{table}

\subsection{Investigation of the actual efficiency of DG-SG DCI}
In addition to the nominal and actual characteristics, the typical average efficiency was estimated in the 1 week experiment during MD-simulations of metal nanocrystals under plastic deformation. The names of tasks (types of jobs), the job runtime ($T_{job}$) on the moderate PC (1 Intel CPU, 2.7 GHz, 2514 MFLOPs, 4GB RAM), the number of jobs ($N_{job}$), the time needed for sequential execution of all jobs on this CPU ($T_{seq}$), the actual total time of parallel execution ($T_{dg}$) for each task (type of the jobs), and speedup $T_{seq}/T_{dg}$ (i.e. the ratio of the total runtime in sequential and distributed modes) are given in the Table~\ref{Table02}. For example, for the task ``S=16x16x16, V=1'' (where $T_{job}=00{^\texttt{h}}15'$) it takes $T_{seq}\approx3.5$ days to calculate 210 jobs \emph{sequentially} on the moderate PC used as the single client. But it takes $T_{dg}\approx0.3$ days only to calculate them \emph{in parallel in DG}, if the same moderate PC used as the DG-server and without any investments in additional hardware and its support. The following notations were used for the names of jobs: S --- the size of the simulated nanocrystal in the lattice spacings (1 spacing for aluminum = 0.4049~nm), and V --- the strain rate (in spacings per second). It should be noted that the subtotal value in column $T_{dg}$ was estimated as the maximal value for this column (because all tasks were performed in parallel), and the total value --- as the sum of the subtotal and the last task ``S=16x16x16, V=0.25'' that was executed separately.

\begin{table}[!ht]
  \centering
    \caption{The comparison table for sequential and parallel execution of jobs in the experiment with massive MD simulations of metal nanocrystals.}
    \label{Table02}
    \begin{tabular}{|c|c|c|c|c|c|}
      \hline
      Task (type of job) & $T_{job}$       & $N_{job}$ & $T_{seq}$ & $T_{dg}$  & Speedup,  \\
                         & (hours:minutes) &           & (days)    & (days)    & $T_{seq}/T_{dg}$ \\ \hline
      \multicolumn{6}{|c|}{Simultaneously executed tasks (all DCI resources are shared among all tasks)}    \\ \hline
      S=16x16x16, V=1    & $00{^\texttt{h}}:15'$      & 210       & 3.5       & 0.3       & 12.1    \\ \hline
      S=16x16x16, V=0.5  & $00{^\texttt{h}}:41'$      & 419       & 11.6      & 6.0       & 1.9     \\ \hline
      S=16x32x16, V=2    & $00{^\texttt{h}}:20'$      & 750       & 10.4      & 3.0       & 3.5     \\ \hline
      S=16x128x16, V=1   & $03{^\texttt{h}}:30'$  & 308       & 44.9      & 5.0       & 9.0     \\ \hline
      S=32x32x32, V=1    & $03{^\texttt{h}}:30'$  & 329       & 48.0      & 4.3       & 11.3    \\ \hline
      S=32x64x32, V=4    & $01{^\texttt{h}}:38'$  & 394       & 27.6      & 6.4       & 4.3     \\ \hline
      S=64x16x64, V=1    & $06{^\texttt{h}}:38'$  & 449       & 125.0     & 4.0       & 31.2    \\ \hline
      \emph{Subtotal}  &         & \emph{2859} & \emph{271.1}  & \emph{6.4} & \emph{42.3} \\ \hline
      \multicolumn{6}{|c|}{Separately executed task (all DCI resources are dedicated to the single task)}     \\ \hline
      S=16x16x16, V=0.25 & 4 hours     & 501       & 83.5      & 1.9       & 43.9    \\ \hline
      \textbf{TOTAL}   &         & \textbf{3360} & \textbf{354.5} & \textbf{8.3}    & \textbf{42.7}    \\
      \hline
    \end{tabular}
\end{table}

\begin{figure}[!ht]
        \begin{center}
            \includegraphics[width=10cm,angle=0]{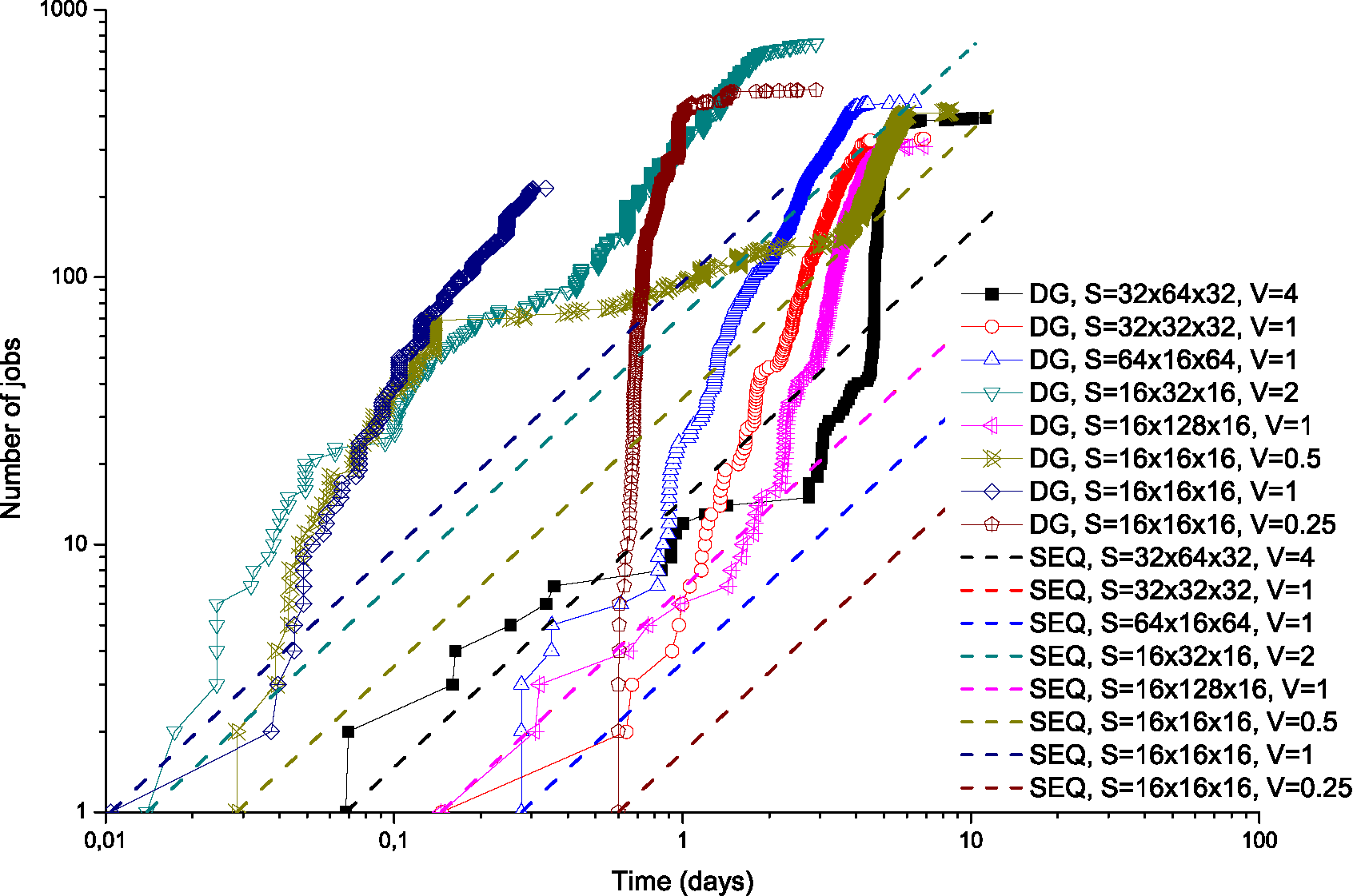}
        \end{center}
        \caption{The number of calculated jobs as a function of time in the parallel distributed (DG) mode (symbols) and sequential (SEQ) mode (dash lines). The different types of jobs are denoted by different symbols and dash lines in the legend. S means the sizes of the simulated nanocrystal (in the lattice spacings), and V means the strain rate (in spacings per second).}\label{Fig05_Jobs_vs_Time}
    \end{figure}

\begin{figure}[hptb]
        \begin{center}
            \includegraphics[width=10cm,angle=0]{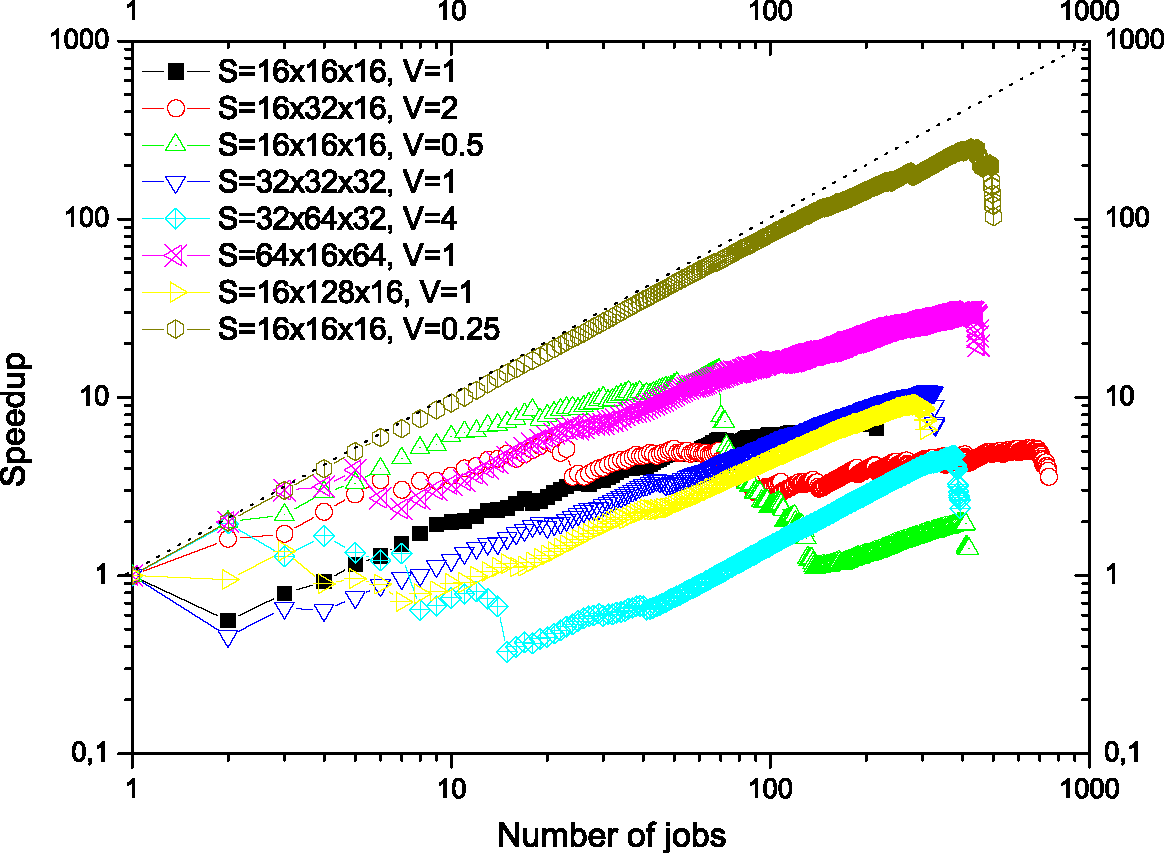}
        \end{center}
        \caption{The effective speedup (the ratio of the total runtime in sequential and parallel distributed modes) for each type of job as a function of the jobs calculated in the parallel distributed (DG) mode (symbols) in comparison with the theoretical maximum (dash line). The different types of jobs are denoted by different symbols in the legend. S means the sizes of the simulated nanocrystal (in the lattice spacings), and V means the strain rate (in spacings per second).}\label{Fig06_Speedup_vs_Jobs}
    \end{figure}

The actual computing rate of these jobs is shown in Fig.~\ref{Fig05_Jobs_vs_Time}, where the 3 distinctive regimes can be found:
\begin{itemize}
  \item ``the initial stage'', when the jobs are mainly distributed among hosts, the number of in-work jobs increases, and the small number of results can be obtained by the master-side;
  \item ``the active stage'', when the number of the finished jobs grows and the effective speedup increases;
  \item ``the final stage'', when no new jobs distributed, the number of the in-work jobs decreases, and the effective speedup decreases.
\end{itemize}
In addition, the everyday (and every hour even) stochastic behavior of plugged/unplugged hosts create the fluctuating behavior of all performance parameters.
That is why the actual speedup is not constant value and can change in the course of the experiment and can be different for the different types of jobs with various demands to the computing resources (Fig.~\ref{Fig06_Speedup_vs_Jobs}). In addition, the several types of jobs those were started simultaneously or with overlapping in the time must compete for the available resources (hosts) and that is why their effective speedups can decrease significantly. Nevertheless, if the number of available resources is higher than the number of the distributed jobs, than the actual speedup (during ``the active stage'') will tend to be closer to the theoretical value. Indeed, the speedup curves (Fig.~\ref{Fig06_Speedup_vs_Jobs}) during ``the active stages'' tend to be parallel to the theoretical level (dash line), i.e. the actual speedup increase with the scaled-up number of jobs is nearly equal to the theoretical limit. It is especially evident for the highest (olive) curve for the type of jobs ``S=16x16x16, V=0.25'', which was calculated alone without any concurrency with other types of jobs.
As a result, the practical feasibility and high efficiency of MD simulations of metal nanocrystals on the basis of DG-SG DCI were demonstrated during this experiment. Finally, it is shown that the total runtime can be decreased by more than 40 times, the whole cost of the similar hardware (at least 189 hosts, i.e. the number of hosts necessary to get such speedup), and their costs of ownership (power supply, support, operation, etc) can be decreased by more than 180 times. In addition, to these achievements, the several important fundamental results were obtained, that can provide the way for the more effective ways of MD simulations in material science and solid state physics.

\section{Results of MD computer simulations in the context of materials science}
From a physical point of view the work was motivated by the previous results
\cite{gordienko1994single,gordienko2006two,kuznetsov2009formation}
that single crystal aluminum foil under the influence of compressed cyclic stretch revealed macro- and micro- scale evolution of crystalline defects in bulk and on surface.
Such defect evolution demonstrates several signs of self-similar geometry and self-organized behavior
\cite{zasimchuk2003equidimensional,gordienko2008multiscale},
that was analyzed by several idealized models
\cite{gordienko1994synergetic,gordienko1996metastable,gordienko1996dynamical,gordienko1995simulation}.
Unfortunately, such models are very rough and do not take into account the peculiarities of interatomic interactions,
especially in distorted crystal lattices.

In this context, the following objectives were formulated for computer simulation:
\begin{itemize}
  \item to simulate several scenarios of plastic deformation in nanocrystals,
  \item to investigate defect accumulation kinetics and evolution of defect substructure;
  \item to find evidences of correlated behavior of individual defects and their complexes;
  \item to measure statistically such correlation in the ensemble of samples with defects.
\end{itemize}

MD simulations of defect substructures
have been performed for several scenarios of the evolution of defect substructure.
The fast parallel computing algorithm \cite{plimpton1995fast} was used for MD simulations of interactions among atoms in the crystalline state
with interatomic potential by Jacobsen, Norskov, Puska (JNP-potential) \cite{jacobsen1987interatomic}
on the basis of embedded atom method (EAM) for aluminum (Al).
DG-enabled application \emph{LAMMPSoverDCI}
was developed to adapt LAMMPS software package for DG-SG DCI \cite{4GLCGW10}.
The calculations were made for Al nanocrystals with $10^5$-$10^7$ atoms
that required the very large scale resources
of the local institutional cluster with 72 CPUs (\texttt{pamela.imp.kiev.ua})
in G.V.Kurdyumov Institute for Metal Physics, National Academy of Sciences of Ukraine (IMP NASU),
``Ukrainian Academic Grid'' with $>10^3$ CPUs, and DG-SG DCI \emph{SLinCA@Home} with $>10^4$ CPUs.

\subsection{Qualitative Results --- Single Nanocrystals}
MD simulations of uniaxial monotonic tensile,
analysis of defects behavior,
and analysis of change of the mode of plastic deformation in Al nanocrystals,
were performed for two orientations: $<$011$>$\{011\} and $<$010$>$\{010\},
where $<$011$>$ and $<$010$>$ denote the direction of the load,
and \{011\} and \{010\} --- planes of test machine grips.
The following figures contain snapshots of the atomic arrangement (with visualization of defects only),
with exclusion of atoms in fcc lattice and atoms at the ends, which were inside the test machine grips and for which the tensile was performed.
They are snapshots of the atom positions near crystalline defects (atoms displaced from the position of the ideal fcc lattice),
that were calculated by the defect determination method based on the common neighbor analysis (CNA)
\cite{tsuzuki2007structural}.
Direction of tensile is from the center to the lower left corner and from the center to the upper right corner,
the rate of loading --- 80~m/s, the number of atoms --- 0.25 million.
The dislocation cores are presented in the form of gray atoms at the edge of the plane of atoms
denoted by dark gray (red in the electronic version) colors.
Dislocation core is formed by the group of atoms with undefined lattice type,
obtained by splitting the total dislocation into two partial dislocations with the formation of stacking faults (SF) in the form of intrinsic and extrinsic SFs, i.e. the planes of atoms in a hcp lattice denoted by dark gray (red in the electronic version) colors.
In addition, some randomly placed perturbations in the form of point-like defects (like atom-vacancy states)
can be observed in groups of 6 neighboring atoms (grey color).

\begin{figure}[hptb]
        \begin{center}
            \includegraphics[width=10cm,angle=0]{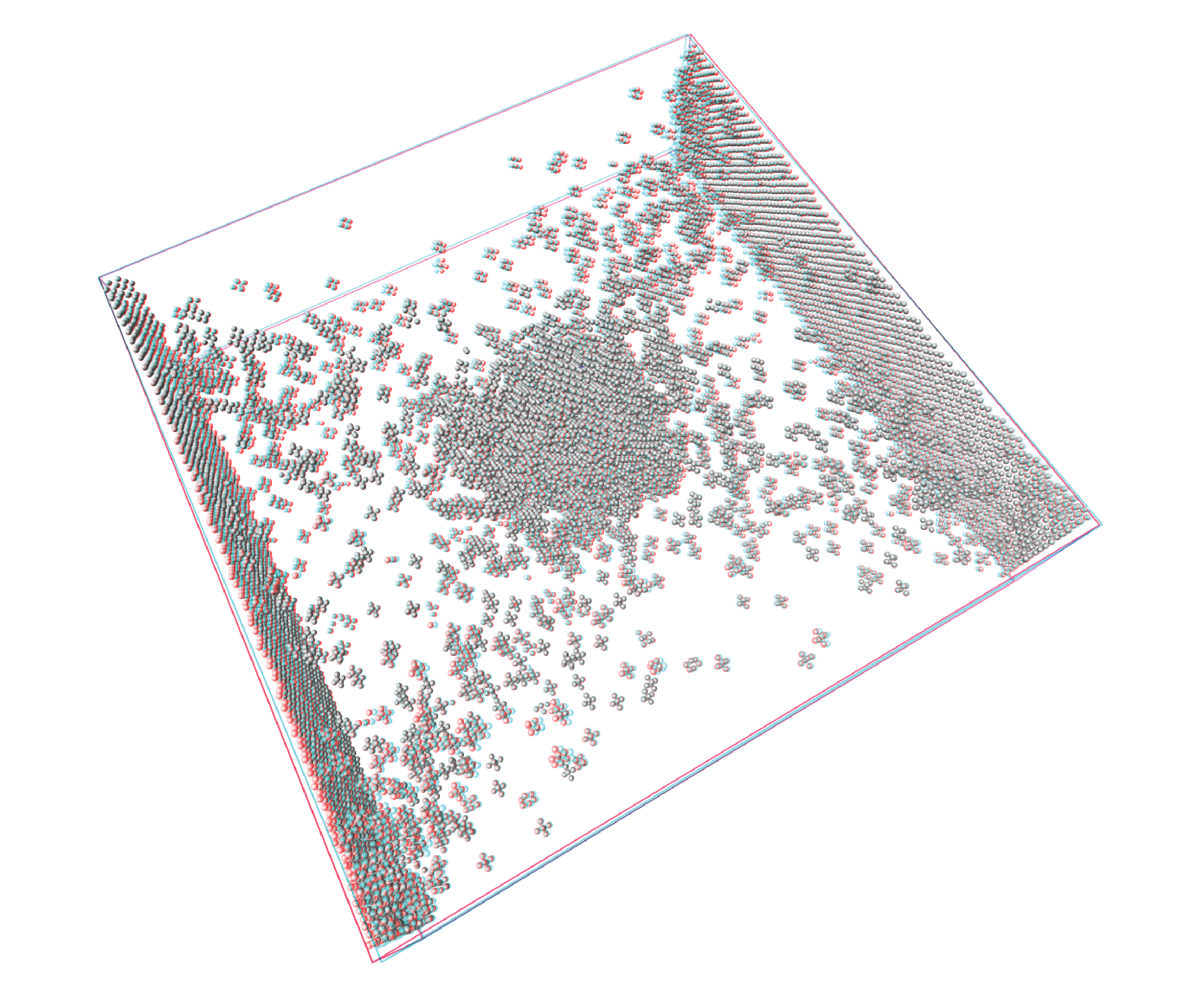}
        \end{center}
                \caption{Tension of Al nanocrystal, strain~$\varepsilon~=~7\%$, orientation $<$010$>$\{010\}.
                (It is anaglyph stereo figure, please, see explanations in the text.)}\label{Fig07_strain7_010}
    \end{figure}

\begin{figure}[hptb]
        \begin{center}
            \includegraphics[width=10cm,angle=0]{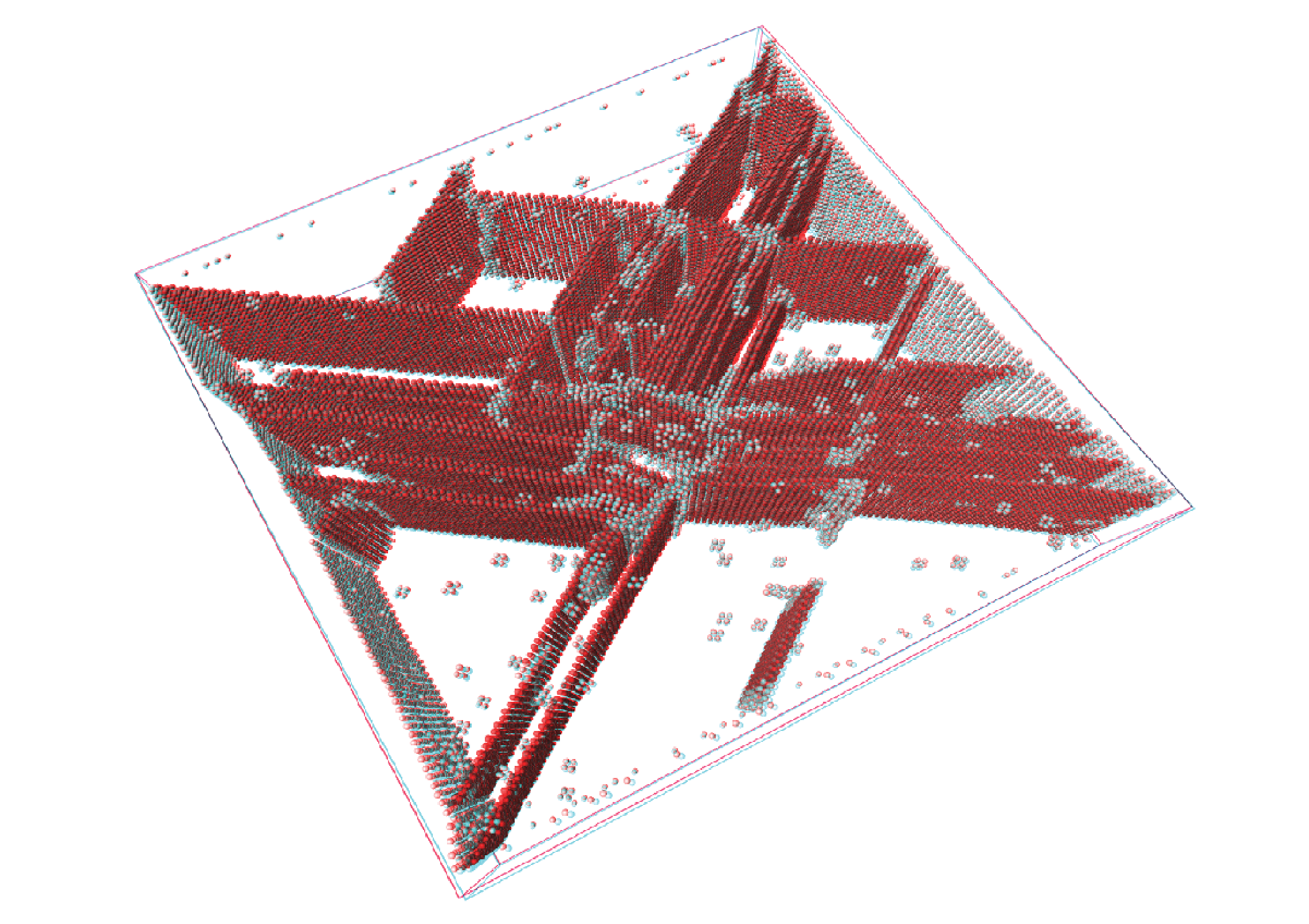}
        \end{center}
                \caption{Tension of Al nanocrystal, strain~$\varepsilon~=~7\%$, orientation $<$011$>$\{011\}.
                (It is anaglyph stereo figure, please, see explanations in the text.)}\label{Fig08_strain7_011}
    \end{figure}

For $<$010$>$\{010\} orientation (Fig.\ref{Fig07_strain7_010})
deformation is accompanied by a growing number of perturbations of the atom-vacancy type,
due to the impossibility of easy glide of dislocations due to the activation the four slip systems with two directions of slip in each with large values of Schmid factor.
(\emph{Note:} This and other anaglyph stereo figures give 3D visual representations of defect substructures,
if anaglyph red-green glasses are used for viewing the color electronic version of the paper.)
For $<$011$>$\{011\} orientation (Fig.~\ref{Fig08_strain7_011})
many intersections of dislocations and SFs are shown that prevent the further easy glide of dislocations.
Further deformation takes place mostly due to the restructuring of existing defect substructures
by correlated displacement of large point-like defect clusters at intersections of SFs and dislocations.
Such behavior is the evidence of the beginning of intense plastic deformation --- collective (hydrodynamic) modes of deformation
(the details can be found in \cite{gordienko2011MDUA}).

\subsection{From Quantity to New Quality --- Statistical Analysis for an Ensemble of Nanocrystals}
\begin{figure}[!ht]
        \begin{center}
            \includegraphics[width=12cm]{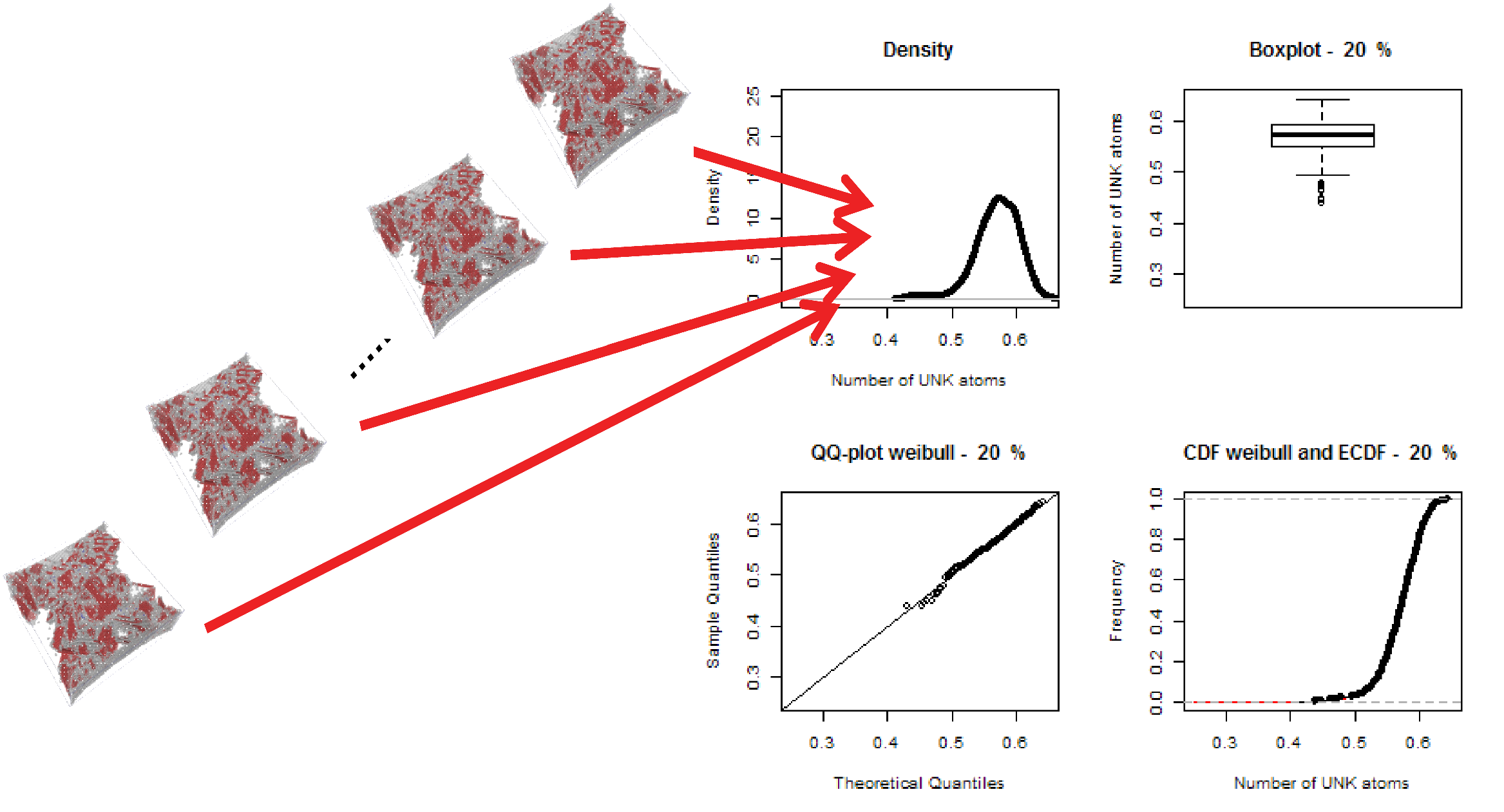}
        \end{center}
          \caption[PDF]
          {Probability density distribution plots (``Density''), boxplots, experimental cumulative density distribution plots (``ECDF'') fitted to Weibull distributions in QQ-plots and CDF plots for defect concentrations in atom-vacancy states (UNK) $\varepsilon=20\%$.}
          \label{Fig09_PDF_CDF}
    \end{figure}

\begin{figure}[!ht]
        \begin{center}
            \includegraphics[width=7cm]{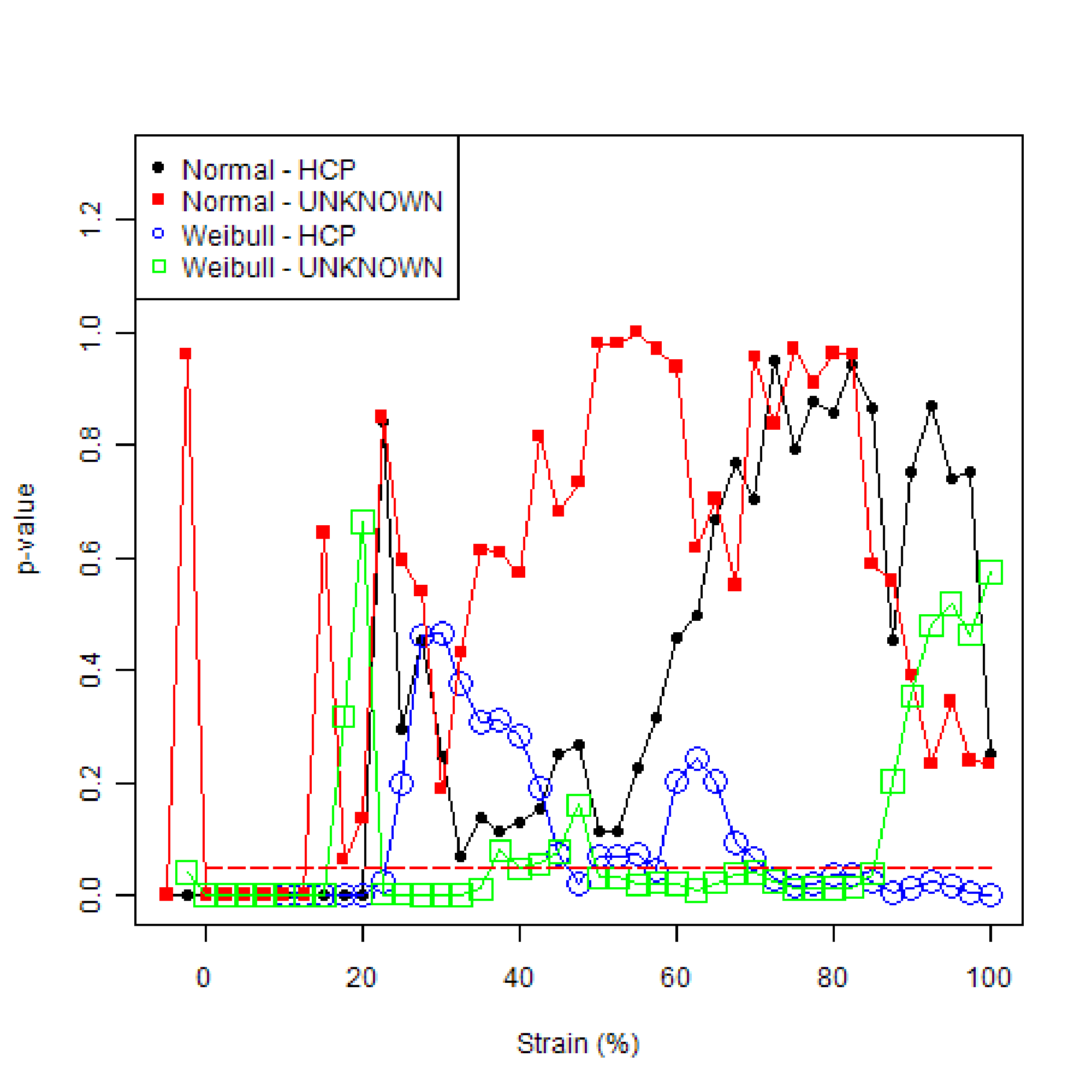}
        \end{center}
        \caption[KS-test]
        {Kolmogorov-Smirnov test for distributions of defect concentrations: SFs (HCP), atom-vacancy states (UNK).}
        \label{Fig10_KS-test}
    \end{figure}

The results of MD simulations for
plastic deformation of the big number of statistical realizations
(i.e. many samples with identical conditions of strain, but different initial random values of atomic velocities) were analyzed.
Distributions of some variables (stresses, concentrations of defects) were evaluated on the basis of the theory of extreme values
\cite{gumbel2004statistics}.
It corresponds to the view on the localized plastic deformation,
as a critical process with correlated behavior of some parts of the deformed crystal
\cite{rinaldi2008peralta}.
The localized plastic deformation can take place
in the areas linked by deformation events (``links'' of the ``chain'' of localized strain),
according to the model of ``a chain with a weak link''.
In such case, the stress values and the defect concentrations for different statistical realizations of the process
(i.e. their distributions over an ensemble of different samples) should be characterized
by the extreme values distributions of I, II, or III type
\cite{rinne2009weibull}.
The aim of the original approach was to carry out MD simulations of plastic deformation under the same conditions (except for the initial random configuration of atomic velocities) for extra-large number of samples (from 400 to 1000 statistical realizations for different deformation conditions),
and to analyze parameter distributions over the ensemble of statistical realizations.
In the context of the current state of the MD simulation of plastic deformation, this approach is essentially new and has no analogues,
and, moreover, it requires very large computing resources.
In Fig.~\ref{Fig09_PDF_CDF} an example of the statistical distribution
over the statistical ensemble of 583 nanocrystals (Al single crystals with $1.5\cdot10^5$ atoms deformed with strain rate 200~m/s in $<$010$>$\{010\} orientation) is shown for the concentration of atoms in atom-vacancy states with the unknown lattice type (UNK)
for $\varepsilon=20\%$.
The results of fitting to Weibull distribution for $\varepsilon=20\%$ (Fig.~\ref{Fig09_PDF_CDF}) give the visually good correspondence to Weibull distribution. Thus, due to these massive character of MD simulations (obtained by means of the DG-SG DCI)
reliability of distribution fitting was possible to estimate by Kolmogorov-Smirnov test (KS-test)
\cite{kolmogorov1933sulla,smirnov1948table}.
The higher values of p-value in the KS-tests mean the higher probabilities of wrong rejection of the current fitting hypothesis,
i.e. the higher probability for the current hypothesis to be at least as extreme as the one that was actually observed.
But the p-value$>0.05$ (the typical significance level)
does not mean that the hypothesis is absolutely true.
Roughly speaking, the higher p-value for the hypothesis of the Weibull distribution,
than for the p-value for the hypothesis of the normal distribution,
means the higher probability of wrong rejection of the former hypothesis
that the estimated distribution follows the Weibull distribution than the normal one.
The results of the KS-tests are shown in Fig.~\ref{Fig10_KS-test}
for the distributions of concentrations of atoms in the hcp lattice (HCP),
i.e. stacking faults (SFs), and concentrations of atom-vacancy states with the unknown lattice type (UNK).
At different stages of deformation (for various values of $\varepsilon$) distribution of the concentrations for defects of both types
can be classified by normal or Weibull distribution, or cannot be distinguished (the details can be found in \cite{gordienko2011MDUA}).
When distributions of defect concentrations with off-center arrangement of the atoms (with an indefinite lattice type) (UNK)
correspond to the Weibull distribution, atom-vacancy states
and their aggregations (see example in Fig.\ref{Fig07_strain7_010}) are assumed to be
correlated and linked by the ``chain'' of strain localization zones
with a power-law distribution of the basic defect concentration over the links in the chain,
according to the theory of extremal values
\cite{gumbel2004statistics,rinne2009weibull}.

\begin{figure}[!ht]
        \begin{center}
        \includegraphics[width=12cm]{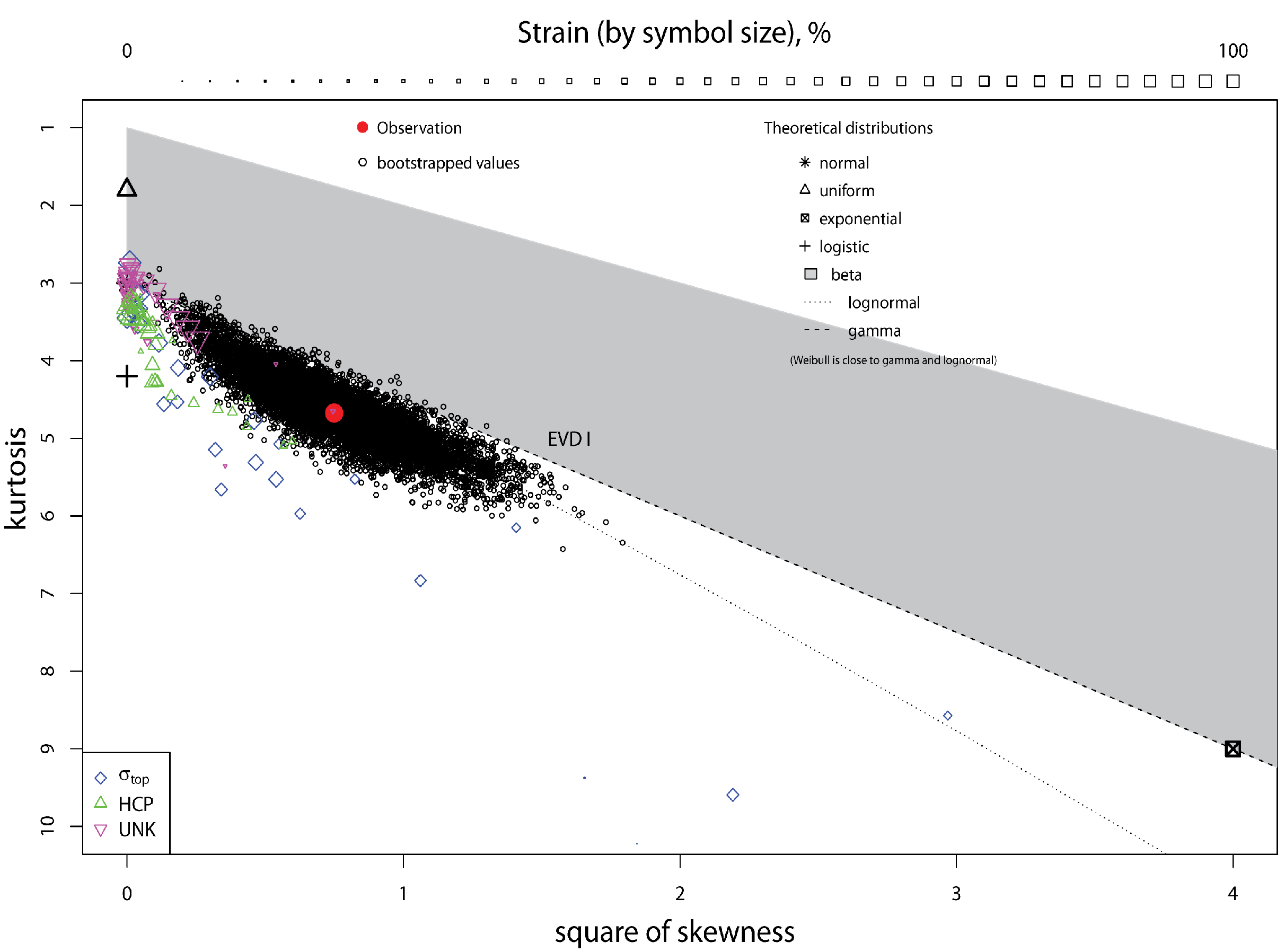}
        \end{center}
        \caption[Moment and bootstrapping analysis]
        {Pearson diagram for moment and bootstrapping analysis for: defect concentration distributions over the ensemble of nanocrystals for SFs (HCP), atom-vacancy states (UNK), and for the normal stress values $\sigma_{top}$ at the atomic planes near grips of the testing machine.}
        \label{Fig11_Bootstrapping}
    \end{figure}

The data of moment and bootstrapping analysis on the Pearson diagram
\cite{cramer1999mathematical,cullen1999probabilistic}
(Fig.~\ref{Fig11_Bootstrapping})
confirm the qualitative results about change of the behavior of the defect substructure from the unrelated state (from the normal distribution)
to the correlated and linked  one (to the zone of Weibull distributions),
and about the change of the deformation mode:
from uncorrelated motion of defects (the normal distribution of defect concentrations and stresses)
to correlated plastic flow (the zone of Weibull distributions).
This indicates availability of the linked local zones (parts of ``chain'') in nanocrystals with hydrodynamic plastic flow
that occurs as the collective motion of defects and their aggregates (atom-vacancy states or SFs)
in the weak link with the power law (scale-invariant or fractal)
basic distribution of the defect concentration over the links of the chain
\cite{gumbel2004statistics,rinne2009weibull}.
As a result of this study, the change of the plastic deformation mechanism was revealed
in Al nanocrystals with $<$011$>$\{011\} and $<$010$>$\{010\} orientations under conditions of localized plastic deformation:
from the uniform laminar flow (based on easy dislocation slip) to the inhomogeneous localized (hydrodynamic) mechanism (based on the correlated movements of groups of atoms, or highly excited atom-vacancy states),
which is in good correlation with our previous experimental studies
\cite{gordienko1994single,gordienko2003unconventional,gordienko2006two}.

\section{Conclusions}
By means of the new HPC technology on the basis of DG-SG DCI the massive MD simulations of plastic deformation processes were carried out for the large quantity of Al nanocrystals ($\sim10^2$-$10^3$).

In the context of computer science, some limitations (fluctuating performance, unpredictable availability of resources, etc.) of the typical DG-SG DCI are outlined, and some advantages (high efficiency, high speedup, and low cost) are demonstrated. Some useful characteristics are given that provide the ways for estimation of the actual metrics of the working DG-SG DCI, for example the normal distribution of host performances (in FLOPs), and signs of log-normal distributions of some other characteristics (CPUs, RAM, and HDD per host) due to the stochastic attachment/detachment of the hosts to the DG-SG DCI with their additive (normal distribution) and multiplicative (log-normal distribution) growth. The practical feasibility and high efficiency of the MD simulations on the basis of DG-SG DCI were demonstrated during the experiment with MD simulations of metal nanocrytstals. Finally, it is shown that the total runtime can be decreased by more than 40 times, the whole cost of the similar hardware, and their costs of ownership can be decreased by more than 180 times in the low-scale DG-SG DCI like \emph{SLinCA@Home} even. The previous estimations show that the big perspectives can be foreseen for combination of the DCI with WS-PGRADE platform \cite{kacsuk2005multi}, that can hugely increase efficiency of scientific computations and simplify complexities of high-performance computing workflows for ordinary material scientists without the special background in computer science.

In the context of physics, the several important fundamental results were obtained, that can provide the way for the more effective ways of MD simulations material science and solid state physics. As an example, the change of plastic deformation mode was investigated under severe plastic strain: from homogeneous (laminar) mode (on the basis of easy dislocation glide) to heterogeneous localized (hydrodynamic) mode (on the basis of correlated displacement of groups of strongly excited atom-vacancy states). The statistical analysis (by Kolmogorov-Smirnov test, moment analysis, and bootstrapping analysis) of the defect density distribution over the ensemble of nanocrystals had shown that such quantitative change of plastic deformation mode is followed by the qualitative change of defect density distribution type over ensemble of nanocrystals. This suggests that some linked local zones (links of chain according to the ``chain with a weak link'' model) of correlated (hydrodynamic) plastic flow appear in a nanocrystal, which manifests as collective displacements of defects and their aggregates in a ``weak link'', with power-law (scale-invariant or fractal) basic distribution of defect density over links of the chain.

\section{Acknowledgments}
  The work presented here was partially funded by EU FP7 DEGISCO (Desktop Grids for International Scientific Collaboration) project, agreement no. RI-261561, EU FP7 SCI-BUS (SCIentific gateway Based User Support) project, contract no. RI-283481;
  and partially supported in the framework of the research theme ``Introduction and Use of Grid Technology in Scientific Research of IMP NASU'' under the State Targeted Scientific and Technical Program to Implement Grid Technology in 2009-2013.



\begin{thebibliography}{10}

\bibitem{4GLCGW10}
Baskova O., Gatsenko O., Gordienko Y.:.
\newblock Enabling high-performance distributed computing to e-science by
  integration of 4th generation language environments with desktop grid
  architecture and convergence with global computing grid.
\newblock In {\it Proc. of Cracow Grid Workshop (CGW'10)}, pages 234--243,
  Cracow, Poland, 2011.

\bibitem{cappello2005computing}
Cappello F., Djilali S., Fedak G., Herault T., Magniette F., Neri V.,
  Lodygensky O.:.
\newblock Computing on large-scale distributed systems: Xtremweb architecture,
  programming models, security, tests and convergence with grid.
\newblock {\it Future Generation Computer Systems}, 21(3):417--437, 2005.

\bibitem{cirne2006labs}
Cirne W., Brasileiro F., Andrade N., Costa L., Andrade A., Novaes R., Mowbray
  M.:.
\newblock Labs of the world, unite!!!
\newblock {\it Journal of Grid Computing}, 4(3):225--246, 2006.

\bibitem{cramer1999mathematical}
Cramer H.:.
\newblock {\it Mathematical methods of statistics}, volume~9.
\newblock Princeton Univ Pr, 1999.

\bibitem{cullen1999probabilistic}
Cullen A., Frey H.:.
\newblock {\it Probabilistic techniques in exposure assessment: a handbook for
  dealing with variability and uncertainty in models and inputs}.
\newblock Springer, 1999.

\bibitem{GatsenkoCGW09}
Gatsenko O., Baskova O., Gordienko Y.:.
\newblock Desktop grid computing in materials science lab - example of
  development and execution of application for defect aggregation simulations.
\newblock In {\it Proc. of Cracow Grid Workshop (CGW'09)}, pages 264---273,
  Cracow, Poland, 2010.

\bibitem{gibrat1931les}
Gibrat R.:.
\newblock {\it Les Inegalites economiques}.
\newblock Paris, France, 1931.

\bibitem{gordienko2008multiscale}
Gordienko Y., Kuznetsov P., Zasimchuk E., Gontareva R., Schreiber J., Karbovsky
  V.:.
\newblock Multiscale 2d rectangular and 3d rhombic gratings created by
  self-organization of crystal structure defects under constrained cyclic
  deformation and fracture.
\newblock {\it Materials Science Forum}, 567:421--424, 2008.

\bibitem{gordienko2011MDUA}
Gordienko Y.:.
\newblock Molecular dynamics simulation of defect substructure evolution and
  mechanisms of plastic deformation in aluminum nanocrystals.
\newblock {\it Metallofizika i Noveishie Tekhnologii (in Ukrainian)},
  33(9):1217--1247, 2011.

\bibitem{gordienko2011generalizedIJMPB}
Gordienko Y.:.
\newblock Generalized model of migration-driven aggregate growth-asymptotic
  distributions, power laws and apparent fractality.
\newblock {\it Int. J. Mod. Phys. B}, 26(1), 2012.

\bibitem{gordienko2012multiplicative}
Gordienko Y.:.
\newblock Multiplicative and additive growth of computing resources in
  dictributed computing infrastructure on the basis of desktop and service
  grids.
\newblock {\it Journal of Grid Computing (to be submitted)}, 2012.

\bibitem{gordienko2006two}
Gordienko Y., Gontareva R., Schreiber J., Zasimchuk E., Zasimchuk I.:.
\newblock Two-dimensional rectangular and three-dimensional rhombic grids
  created by self-organization of random nanoextrusions.
\newblock {\it Advanced Engineering Materials}, 8(10):957--960, 2006.

\bibitem{gordienko1994single}
Gordienko Y., Zasimchuk E.:.
\newblock Single-crystal indicators of fatigue and plastic deformation damage.
\newblock {\it Proceedings of SPIE}, 2361:312, 1994.

\bibitem{gordienko1994synergetic}
Gordienko Y., Zasimchuk E.:.
\newblock {Synergetic model of structure formation during plastic deformation
  of crystals}.
\newblock {\it Philosophical Magazine A}, 70(1):99--107, 1994.

\bibitem{gordienko1995simulation}
Gordienko Y., Zasimchuk E.:.
\newblock Simulation of building one-and two-dimensional structures on many
  scales in metals under load.
\newblock {\it Systems Analysis Modelling Simulation}, 18:837--840, 1995.

\bibitem{gordienko1996dynamical}
Gordienko Y., Zasimchuk E.:.
\newblock Dynamical phase transition in a lattice gas model with aggregation
  and self-organization.
\newblock {\it Physica A: Statistical and Theoretical Physics},
  229(3-4):540--551, 1996.

\bibitem{gordienko1996metastable}
Gordienko Y., Zasimchuk E.:.
\newblock Metastable fractal aggregates as a result of competition between
  diffusion-limited aggregation and dissociation.
\newblock In {\it Proc. of the 8th Joint EPS-APS Int. Conf. on Physics
  Computing: PC'96: September 17-21, 1996, Krak{\'o}w, Poland}, page 293.
  Academic Computer Centre CYFRONET-KRAK{\'O}W, 1996.

\bibitem{gordienko2003unconventional}
Gordienko Y., Zasimchuk E., Gontareva R.:.
\newblock Unconventional deformation modes and surface roughness evolution in
  al single crystals under restricted cyclic tension conditions.
\newblock {\it Journal of Materials Science Letters}, 22(3):241--245, 2003.

\bibitem{gumbel2004statistics}
Gumbel E.:.
\newblock {\it Statistics of extremes}.
\newblock Dover Pubns, 2004.

\bibitem{jacobsen1987interatomic}
Jacobsen K., Norskov J., Puska M.:.
\newblock Interatomic interactions in the effective-medium theory.
\newblock {\it Physical Review B}, 35(14):7423, 1987.

\bibitem{Kacsuk2009}
Kacsuk P., Kovacs J., Farkas Z., Marosi A., Gombas G., Balaton Z.:.
\newblock Sztaki desktop grid (szdg): a flexible and scalable desktop grid
  system.
\newblock {\it Journal of Grid Computing}, 7(4):439--461, 2009.

\bibitem{kacsuk2005multi}
Kacsuk P., Sipos G.:.
\newblock Multi-grid, multi-user workflows in the p-grade grid portal.
\newblock {\it Journal of Grid Computing}, 3(3):221--238, 2005.

\bibitem{kolmogorov1933sulla}
Kolmogorov A.:.
\newblock Sulla determinazione empirica di una legge di distribuzione.
\newblock {\it Giornale dell’Istituto Italiano degli Attuari}, 4(1):83--91,
  1933.

\bibitem{kuznetsov2009formation}
Kuznetsov P., Petrakova I., Gordienko Y., Zasimchuk E., Karbovskii V.:.
\newblock Formation of self-similar structures on $\{$100$\}$ 00l aluminum
  single-crystal foils under cyclic tension.
\newblock {\it Physical Mesomechanics}, 12(1-2):85--93, 2009.

\bibitem{plimpton1995fast}
Plimpton S.:.
\newblock Fast parallel algorithms for short-range molecular dynamics.
\newblock {\it Journal of Computational Physics}, 117(1):1--19, 1995.

\bibitem{rinaldi2008peralta}
Rinaldi A., Peralta P., Friesen C., Sieradzki K.:.
\newblock Sample-size effects in the yield behavior of nanocrystalline nickel.
\newblock {\it Acta Materialia}, 56(3):511--517, 2008.

\bibitem{rinne2009weibull}
Rinne H.:.
\newblock {\it The Weibull distribution: a handbook}.
\newblock Chapman \& Hall/CRC, 2009.

\bibitem{smirnov1948table}
Smirnov N.:.
\newblock Table for estimating the goodness of fit of empirical distributions.
\newblock {\it The Annals of Mathematical Statistics}, 19(2):279--281, 1948.

\bibitem{tsuzuki2007structural}
Tsuzuki H., Branicio P., Rino J.:.
\newblock Structural characterization of deformed crystals by analysis of
  common atomic neighborhood.
\newblock {\it Computer physics communications}, 177(6):518--523, 2007.

\bibitem{urbah2009edges}
Urbah E., Kacsuk P., Farkas Z., Fedak G., Kecskemeti G., Lodygensky O., Marosi
  A., Balaton Z., Caillat G., Gombas G., others.:.
\newblock Edges: bridging egee to boinc and xtremweb.
\newblock {\it Journal of Grid Computing}, 7(3):335--354, 2009.

\bibitem{zasimchuk2003equidimensional}
Zasimchuk E., Gordienko Y., Gontareva R., Zasimchuk I.:.
\newblock {Equidimensional fractal maps for indirect estimation of deformation
  damage in nonuniform aircraft alloys}.
\newblock {\it J. Mater. Eng. Perf.}, 12(1):68--76, 2003.

\end{thebibliography}
\end{document}